%% file: these-v2.tex
\documentclass[10pt,a4paper]{./style/book2}

\usepackage[french,english]{babel}
\usepackage[applemac]{inputenc}
\usepackage{cite}
\usepackage{graphicx}
\usepackage{psfrag}
\usepackage{stfloats}
\usepackage{amsmath}
\usepackage{array}
\usepackage{topcapt}
\usepackage[nolineno]{./style/lgrind}
\usepackage{color}
\usepackage{minitoc}
\usepackage{fancyhdr}
\usepackage[Lenny]{./style/fncychap}

\usepackage{graphics}
\usepackage{graphicx}
\usepackage{amssymb}
\usepackage{lineno}
\usepackage{subfigure}

\usepackage{multirow}

\input{./chap_tex/headers.tex}

\begin{document}

\newif\ifnumero          
\newif\ifnumcont         
\newif\ifabstractaudos   
\numerotrue \numconttrue \abstractaudostrue






\input{./chap_tex/pagetitre-en.tex}




\newcommand{\myfancy}{
\pagestyle{fancy}
\fancyhead[LE]{\thepage}
\fancyhead[RE]{\slshape \rightmark}
\fancyhead[RO]{\thepage}
\fancyhead[LO]{\slshape \leftmark}
\cfoot{}
}
\myfancy


\pagestyle{empty}

\begin{center}
\mbox{}\\
{\Large\bf Abstract \\}
\end{center}
In this report, we first describe the problem that we are dealing with i.e. data dissemination in multi-hop cognitive radio networks. To address this problem, we propose a channel selection strategy named `SURF'. We evaluate the proposed channel selection strategy in both single-hop and multi-hop scenarios and compared it with relevant approaches. So far, one technical report and a poster is published as part of this work, while two publications are under review; one is in IEEE Communications Letters and the second one is in IEEE WoWMoM conference. In on-going works sections, we first mention some possible directions in the context of SURF. In addition to that, we mention different research problems that we are planning to deal during the course of this PhD dissertation.

%


\pagestyle{empty}

\begin{center}
\mbox{}\\
{\Large\bf Publications \\}
\end{center}
\vspace{-0.8cm}
\begin{flushleft}
\mbox{}\\
{\Large\bf Papers Under Review \\}
\end{flushleft}

\begin{itemize}
\item Mubashir Husain Rehmani, Aline Carneiro Viana, Hicham Khalife, and Serge Fdida, {\it Channel Assortment Strategy for Reliable Communication in Multi-hop Cognitive Radio Networks}, {\bfseries Submitted to: IEEE Communications Letters}.
\item Mubashir Husain Rehmani, Aline Carneiro Viana, Hicham Khalife, and Serge Fdida, {\it Toward Reliable Contention-aware Data Dissemination
in Multi-hop Cognitive Radio Ad Hoc Networks}, {\bfseries Submitted to: IEEE WoWMoM 2010}.
\end{itemize}

\begin{flushleft}
\mbox{}\\
{\Large\bf Technical Report \\}
\end{flushleft}

\begin{itemize}
\item Mubashir Husain Rehmani, Aline Carneiro Viana, Hicham Khalife, and Serge Fdida, {\it Toward Reliable Contention-aware Data Dissemination
in Multi-hop Cognitive Radio Ad Hoc Networks}, INRIA RR-0375, 2009, France. http://hal.inria.fr/inria-
00441892/en/.
\end{itemize}

\begin{flushleft}
\mbox{}\\
{\Large\bf Poster \\}
\end{flushleft}

\begin{itemize}
\item Mubashir Husain Rehmani, Aline Carneiro Viana, Hicham Khalife, and Serge Fdida, {\it Adaptive and Occupancy-based Channel Selection for unreliable Cognitive Radio Networks}, Rencontres Francophones sur les Aspects Algorithmiques des Telecommunications (ALGOTEL) 2009, du 16 au 19 juin 2009, Carry Le Rouet, France.
\end{itemize}

\begin{flushleft}
\mbox{}\\
{\Large\bf Talks \\}
\end{flushleft}

\begin{itemize}
\item Mubashir Husain Rehmani, Aline Carneiro Viana, Hicham Khalife, and Serge Fdida, {\it Toward Reliable Contention-aware Data Dissemination
in Multi-hop Cognitive Radio Ad Hoc Networks}, Presented in the workshop "Groupe de Travail" for the students of Masters in Computer Science, Speciality Networks (RES), University Pierre and Marie Curie (UPMC), Jussieu, Paris, France, February 2010.
\end{itemize}


\baselineskip=16pt 

\begin{center}
\mbox{}\\
{\Huge\bf Data Dissemination in Cognitive Radio Networks \\}
\end{center}

%
%
%







\pagestyle{plain}
\setcounter{page}{1}
\setcounter{chapter}{1}
\section{Introduction}

Radio spectrum is a precious natural resource and it has been utilized for communication since many decades. In this perspective, several techniques have been proposed to utilize the radio spectrum efficiently. These techniques ranges from geographical reutilization of radio spectrum and multiplexing techniques, just to name a few. Nevertheless, Federal Communication Communications (FCC) reports that the radio spectrum is still under-utilized and the occupancy of radio spectrum varies from $15\%-85\%$~\cite{survey}. According to FCC, this is due to fixed spectrum assignment policy and geographical and temporal utilization of the radio spectrum. Thus, keeping this notion in mind, the idea of Cognitive Radio Ad-Hoc Networks (CRN) is first coined by J. Mitola {\it et al.}~\cite{mitola} to utilize the radio spectrum efficiently. In fact, Cognitive Radio Networks are composed of {\it intelligent} wireless nodes a.k.a. Cognitive Radio (CR) nodes. These CR nodes scan the spectrum and utilizes the spectrum as soon as they found the spectrum is free by the legacy users a.k.a. Primary Radio (PR) nodes.


Cognitive radio ad-hoc networks have been widely used in Military and Emergency Networks due to their mission critical nature~\cite{militarysdr},~\cite{ossama}. But with the advancement in technology and the availability of cheap consumer devices, the applications of cognitive radio networks for consumer-based applications is no more longer a dream. In this perspective,~\cite{applications},~\cite{buddhikot} and~\cite{chapin} highlights some of the consumer-based applications of cognitive radio networks. These types of consumer-based cognitive radio network applications can be envisaged in near future~\cite{applications} and can be deployed in Railway stations, markets, airports, public spots, and restaurants. One prospective application could be data dissemination in these types of networks which enables a CR user to convey emergency alarms and alerts or to deliver low
priority data such as advertisement messages in a cognitive radio multi-hop context.


Data dissemination is a classical and a fundamental function in any kind of network. In wireless
networks, the characteristics and problems intrinsic to the wireless links bring several challenges in data
dissemination in the shape of message losses, collisions, and broadcast storm problem, just to name a few.
However, in the context of Cognitive Radio Ad-Hoc Network (CRN)~\cite{survey}, reliable data dissemination is much
more challenging than traditional wireless networks.
First, in addition to the already known issues of wireless environments,
the diversity in the number of channels each cognitive node can use adds another
challenge by limiting node's accessibility to its neighbors. Second, cognitive
radio nodes have to compete with the Primary Radio (PR) nodes for the residual resources on many channels and
use them opportunistically. Besides, during communication CR nodes should communicate in such a way that
it should not degrade the reception quality of PR nodes by causing CR-to-PR interference. In addition, CR
nodes should immediately interrupt its transmission whenever a neighboring PR activity is
detected~\cite{hicham}.

\begin{figure}[h]
    \begin{center}
    \includegraphics[width=9.5cm]{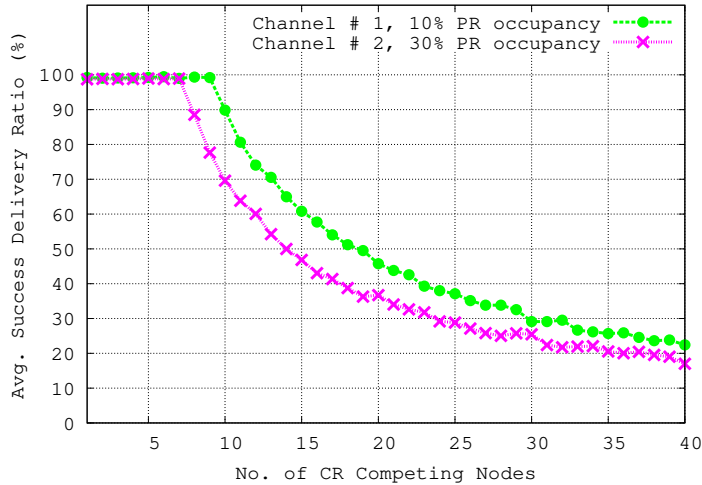}
\caption{CR nodes competing for the same channel.}
    \label{fig1-a}
\end{center}
\end{figure}

\vspace{-0.5cm}

Recently, a lot of work has been carried out for dynamic channel management in cognitive radio 
networks~\cite{cao,pomdp,niyato,rahul}. However, all these approaches focuses on single-hop 
cognitive radio networks and either requires the presence of any central entity or coordination with primary
radio nodes in their channel selection decision. For instance,~\cite{cao} proposed an efficient spectrum 
allocation architecture that adapts to dynamic traffic demands but they considered a single-hop scenario of 
Access Points (APs) in Wi-Fi networks.~\cite{mubashir} proposed a channel selection strategy based on the primary 
user's occupancy but specifically designed for single-hop architecture. 

In multi-hop cognitive radio ad hoc networks, where coordination between CRs is hard to achieve and 
no central entity for regulating the access over channels is to be envisaged,
reliable data dissemination is even more complex.
In this perspective, the first step in
having efficient data dissemination is  to know {\it how to select best channels}. Thus, differently from
most works in the literature dealing with single-hop communication~\cite{cao},~\cite{jcm}, we go a step further
here and build up a channel selection strategy for multi-hop communication in CRN.
The objective of every cognitive radio node is to select the best channel ensuring a maximum
connectivity and consequently, allowing the largest data dissemination in network. This corresponds to
the use of channels having not only low primary radio nodes (PRs) activities, nevertheless the reliability
of the dissemination process is achieved by limiting the contention of
cognitive ratio nodes (CRs) acceding selected channels.

The effect of CR contentions on dissemination is highlighted in Fig.~\ref{fig1-a} that shows the evolution of the average
success delivery ratio at receivers of a single
source, with the number of competing CRs.
It is clear that the performance of a channel with low PR activity
decreases with the number of CR competing for the available resource. Nevertheless, a channel with higher PR activity can be a
good choice if CR contention is low. The challenge here is then how to find a good tradeoff
between connectivity and contention.

In this report, we present our work i.e. a channel selection strategy, named SURF.
The goal of SURF is to ensure reliable
contention-aware data dissemination and is specifically designed for multi-hop cognitive radio ad hoc
networks. Usually channel selection strategies provide a way to nodes to select channels for transmission.
On the contrary, SURF endue CR nodes to select best channels not only for transmission but also for
overhearing. As a result, both sender and receiver tuned to the right channel for effective and reliable data
dissemination. Additionally, by dynamically exploring residual resources on channels and by monitoring the
number of CRs on a particular channel, SURF allows building a connected network with limited contention where
reliable communication can take place. 

The remainder of this report is organized as follows: we
discuss connectivity vs. contention trade-off in Section~\ref{motiv}. We
give general overview of our channel selection strategy SURF in Section~\ref{overview}. Section~\ref{dd} deals with performance evaluation of SURF. Finally, section~\ref{futureworks}
briefly describe the on-going works and section~\ref{conclusion} concludes the report. 

\section{Cognitive radio ad hoc networks: connectivity VS. contention trade-off}
\label{motiv}

In a highly dynamic/opportunistic cognitive radio network, cognitive users compete for residual
resources (a.k.a spectrum holes) left by the activity
of the legacy users more formally called primary radio users. Every cognitive node, using an intelligent selection strategy, selects the appropriate
channel for transmitting with the major constraint of not degrading the service of ongoing primary radio communications. Indeed,
primary radios have the absolute priority over the communication channels.
In an opportunistic multi-hop cognitive radio network where coordination between CRs is hard to achieve and no central entity for regulating
the access over channels is to be envisaged, the objective of every cognitive radio is to select the channel ensuring a maximum connectivity.
Such spectrum band has the highest number of active cognitive radios hence allows quick and effective data dissemination in the network.

\begin{figure}[h]
\begin{center}
    \includegraphics[width=14cm]{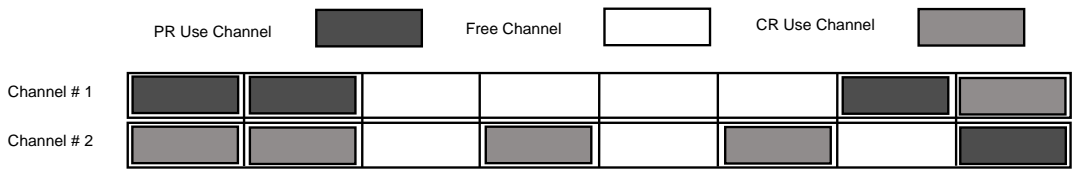}
\caption{PR and CR Nodes occupancy over channels}
    \label{fig1-b}
    \end{center}
\end{figure}

Intuitively, one may think that the best strategy for \emph{all} CRs is to dynamically switch to the less
occupied channel (by PRs). Thus, satisfying the objective of verifying priority constraints imposed by PRs.
Nevertheless, such strategy leads to
many classical problems already well known in wireless networking. First, forcing all CRs in a geographic
area to be active over the same channel makes all nodes compete for the same resource thus generating
contention and collision problems. Second, such approach wastes the valuable additional capacity on different
channels that the cognitive radio concept offers. Indeed, it was already shown in traditional wireless
networking that networks with high contention, where repetitive collisions are frequent, suffer from
close to zero throughput~\cite{jingyang}. A typical example is described in Fig.~\ref{fig1-b}.
Initially, channel 1 has more primary radio activities and should be avoided by the CR transmitters. However,
if enough CRs switch to channel 2 to communicate, channel 1 quickly becomes less occupied and able to carry
higher throughputs than channel 2. Therefore, taking into account contention issues due to CR transmissions is
necessary when selecting spectrum bands for CR communications.

\vspace{-0.8cm}Any proposed strategy for channel selection in CRN has to optimize the connectivity vs. contention trade-off.
We propose hereafter a channel selection strategy that monitors the number of active CR nodes on a particular
channel. As a result, {\it we are able to build a well connected network while dynamically exploiting
residual resources on many channels}. 

\section{SURF: Channel Selection Strategy}
\label{overview}

SURF channel selection strategy is specifically designed for ad hoc cognitive radio networks.
The general goal of our strategy is to ensure a reliable data dissemination over a multi-hop CRN. Such technique can
be used to convey emergency alarms and alerts or to deliver low priority data such as advertisement messages
in a cognitive radio multi-hop
context. Recall that in order to achieve our goal
and ensure coverage and reliability, the connectivity vs. contention trade-off should be optimized.

SURF strategy is exclusively implemented by every CR node and is used for transmission and/or overhearing.
Using the decentralized algorithm proposed by SURF, every CR sender judiciously selects the \textit{best}
frequency band for sending messages and every CR receiver tunes to the right channel (selected by the sender)
to retrieve the sent data.

With SURF, each CR node looks first for the less PR-occupied
channel to help deciding autonomously which channel to use.
In addition to PR occupancy, we also consider CR neighbors competing for the same channel resource.
More precisely, every CR node classifies available channels based on the observed PR-occupancy over these
channels. This classification is then refined by identifying the number of active CRs over each band. The
best channel for transmission is the channel that has the lowest PR activity and a reasonable ongoing CR activity.
Indeed, choosing a channel with few CRs yields to a disconnected network. The challenge in our strategy is in
finding the number of active CRs on every channel that gives the best connectivity with limited contention.
Practically, every CR after classifying available channels, switches dynamically to the best one and broadcasts the stored message.

Additionally, CRs with no messages to transmit implement the SURF strategy in order to tune to
the \emph{best} channel for data reception. Clearly, using the same strategy implemented by the sender allows
nodes in the close geographic areas to select the same channel as sender for overhearing with high probability.
Intuitively, it is likely that CRs in the sender's vicinity have the same PR occupancy, hence channels available
to a CR sender is also available to its neighbors with high probability~\cite{pomdp}.
Therefore, SURF controls the number of CR receivers, thus a connected topology with low contention is created.
Once a packet is received, every CR receiver undergoes again the same procedure to choose the appropriate
channel for conveying the message for its neighbor.

Through simulations\footnote{For more detailed simulations results please refer our Technical Report~\cite{tr}}, we show that SURF builds, as expected, a highly connected network
suitable for reliable dissemination.
Moreover, SURF outperforms existing algorithms.
In order to evaluate the performance of SURF, we compare it with an intuitive random
strategy (RD) and the two variants of selective broadcasting protocol~\cite{agrawal}, i.e. selective
broadcasting strategy (SB)~\cite{agrawal} without any centralized authority, and selective broadcasting
with centralized authority (CA) to be served as an upper bound.
The simplicity and decentralized nature of our
solution makes it usable in ad hoc CRNs deployed to convey emergency messages and alerts. It can also be employed in
commercial applications to disseminate short publicity messages.

\section{Performance Evaluation}
\label{dd}

To assess the performance of SURF with RD, SB, and CA in term of reliable data dissemination,
two performance metrics are evaluated with different total number of channels:
(i) the average delivery ratio, which is the ratio of packet received by a
particular CR node over total packets sent in the network and (ii) the average number of accumulative CR receivers at
each transmission, until TTL=0. Recall that higher number of channels yields to lower PR occupancy.
In addition, it is worth mentioning here that even the centralized approach CA could not get a 100\% of data
dissemination because of the performed randomly assignment of $Acs$ set to CR nodes. This may generate topology disconnections
caused by physical close nodes being assigned to disjoint channels. In this way, as previously stated, we consider
the CA approach gets the theoretical upper bound results in terms of message dissemination.

\begin{figure*}[!ht]
\begin{center}
\subfigure[Accumulative receivers]{
\includegraphics[width=8.5cm]{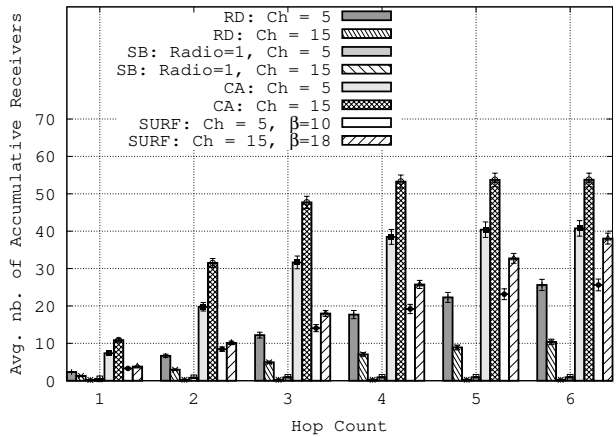}
\label{fig9}
}
\subfigure[Delivery ratio]{
\includegraphics[width=8.5cm]{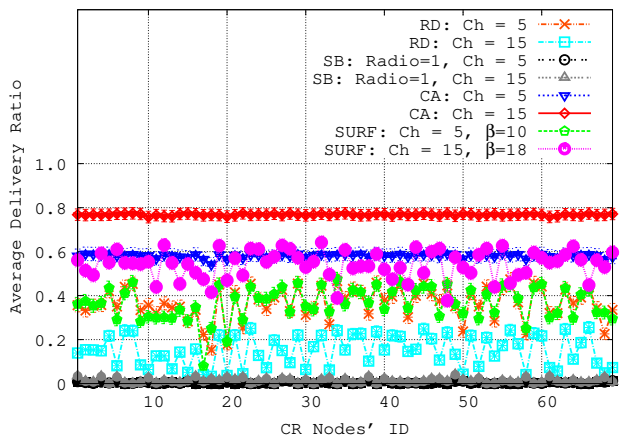}
\label{fig10}
}
\caption{Average number of accumulative receivers per hop and average delivery ratio in a 70-node CRN,
       for random (RD), selective broadcasting (SB), centralized approach (CA), and our strategy (SURF).}
\label{receivers_ratio}
\end{center}

\end{figure*}





Fig.~\ref{fig9} compares the number of accumulative CR receivers at each hop of communication until TTL=0,
for the four strategies. When $Ch=15$, SURF allows the message dissemination to 55\% of nodes in the network (i.e. 38
out of 70 CR nodes), while CA allows 78\% (i.e. 54 over 70 CR nodes).
Additionally, due to its central control and multiple transmissions, the CA strategy reaches this upper bound of receivers
percentage at the TTL=4.
It can be clearly seen that SURF outperforms RD and SB and compared to CA, only provides a decrease of 25\% in performance. The gain
achieved with CA is at the price of more transmissions, more energy consumption, and more expensive and sophisticated devices.

Fig.~\ref{fig10} compares delivery ratio of RD, SB, CA and SURF, as a function of the CR nodes' ID. SURF outperforms RD
and SB in terms of delivery ratio, when number of channels are high. Compared to the CA strategy, SURF has only
20\% of performance reduction. In particular, for {\it Ch=5} and {\it Ch=15}, SURF guarantees the delivery of
approximately 60\% of messages (with a single transmission), contrarily to less
than 20\% for the RD and SB strategies (with single and multiple transmissions, respectively) and
80\% for the CA strategy (with multiple transmissions).


\section{On-Going Works}
\label{futureworks}
In this section, we describe the goals that we want to achieve during the course of PhD Thesis. The on-going works of our thesis encompasses the following themes:

\subsection{Extension of SURF}
Our proposed channel selection strategy `SURF' can be extended in several directions. One possibility is to investigate and optimize data dissemination delay and compared it with other relevant approaches. This delay in data dissemination is due to message losses when sender and receiver are not on the same channel. In addition to that we can add a utility-based heuristic to the channel weight computation and investigates SURF performance under dynamic traffic by consideration of data rates and traffic volume generated by CR nodes in the channel weight formula. By considering the data rates, a channel that supports higher data rates can be selected. While, considering the CR traffic volume avoid congestion in the network. Moreover, prediction and history can also be accounted in SURF to enhance the performance. In fact, if a pattern exists in the traffic utilization of CR nodes, then by the help of prediction techniques, one can estimate the future occupancy of CR nodes over channels and avoid highly CR occupied channels.


\subsection{Study of the NS model for Cognitive Radio Networks}
We are studying the Network Simulator (NS) model for Cognitive Radio Networks. In fact, we found a NS-2 patch~\cite{crcn} available for Cognitive Radio Networks that support many functionalities of Cognitive Radio Network. Thus, our goal is to first study in detail the existing patch of NS Simulator and use this existing patch for our future simulations.


\subsection{Data Dissemination through Channel Bonding in Multi-Hop Cognitive Radio Networks}
Based upon our channel selection strategy SURF, we incorporate another important aspect that can be exploited in the context of Cognitive Radio Network for efficient and reliable communication. In fact, we exploit the availability of contiguous non-overlapping channels to create a bonded channel and use it with our channel selection strategy.

Through channel bonding~\cite{draft,chnbond,chnaggre}, multiple frequency channels are bonded into a single broadband channel. Therefore, the aggregated bandwidth is large due to the sum of multiple frequency channels and as a consequence, the rate of packet transmission increase. This will also reduce the packet transmission time. Another advantage of channel bonding is the low delay.

Thus, we propose a broadcasting strategy based
on Channel Bonding, named `BOND'; empowers CR nodes
with the ability to infer, based on information regarding
PR occupancy, the less PR-occupied channel to use. Once
classified the less PR-occupied channels, CR nodes detect and
group the contiguous non-overlapping channels to create a
bonded channel. Cognitive radio nodes then use this bonded
channel for broadcasting in order to improve the performance
of the network. We claim that if we do not exploit the usage of nonoverlapping
contiguous channels, the spectrum will be used
inefficiently.

In this way, how to find a good transmission opportunity
in terms of PR occupancy and selection of bonded channel,
constitutes our main goal. BOND will provide
to CR nodes a strategy to select channels to create a bonded
channel according to their availabilities, giving to nodes the
possibility of selecting the best classified one for transmission
and/or overhearing.

In other words, the use of channel bonding in conjunction
with our channel selection strategy, allow CR users to efficiently
disseminate and share information. Our strategy is adaptive
in nature and well suited for cognitive radio networks, where
the available channel set changes dynamically. In future, we intend to characterize the number of channel to be bonded and it's impact over network performance metrics. Through simulations,
we intend to demonstrate that our solution is scalable as the
number of channel increases. Note that this is on-going work and we are thinking over several dimensions.


\subsection{Internet Access Framework for Future Cognitive Radio Networks}
We have gained some insights through the work that we have done so far on channel selection strategies, challenges of broadcasting, and data dissemination in cognitive radio networks. Thus, by combining features of our proposed channel selection strategy SURF, channel bonding, and data dissemination, we intend to use them to create a basis to build upon a framework for future cognitive radio networks. This framework is both for challenged environments and can be extended to consumer-based cognitive radio networks applications.

In this context, an Internet Access Framework for future Cognitive Radio Networks (IAFCRN) in under investigation. Through this framework, cognitive radio nodes achieves
the goal of robust connectivity to connect to internet and disseminate data in challenged environments. Moreover, IAFCRN assures efficient utilization of the spectrum usage and
increase the capacity of the network by exploiting channel bonding and channel aggregation techniques. The proposed framework
is completely decentralized in nature by integrating and exploiting the new paradigm of Cognitive Radio Sensor Networks. Thus,
by deploying low-cost sensors in the vicinity of cognitive radio access points, a spectrum opportunity map is created to facilitate
channel selection decision. This feature makes the IAFCRN framework coordination-independent with the existing infrastructure
i.e. Primary Network. Moreover, deploying sensors instead of cognitive radio base station requires less complexity and cost. We
then investigates the applicability of existing solutions and their shortcomings in the context of IAFCRN framework. We envisage
that by integrating channel bonding, channel aggregation, and cognitive radio sensor networks, new consumer-based cognitive
radio communication paradigms can be realized. Validation through simulations are left for future works.


\subsection{Challenges of Broadcasting in Cognitive Radio Networks}
At the beginning of the Thesis, we have done a comprehensive literature review on broadcasting strategies. Now, based on the experience that we have acquired,
we are writing a survey paper on broadcasting strategies in Cognitive Radio Networks. In this survey paper, we give an overview of various broadcasting strategies that have been proposed so far for
Cognitive Radio Networks. Moreover, we also intend to give a comprehensive survey of broadcasting schemes for Cognitive Radio Networks. We identify required key characteristics of broadcasting strategies in Cognitive Radio Networks. In addition to that we classify these schemes and we believe that this classification would be useful for academic and industry based researchers who are engaged in the design of broadcasting strategies for Cognitive Radio Network. We also give some insights to improve these schemes, which may be helpful to researchers who want to further improve them.


\section{Conclusion}
\label{conclusion}
For efficient spectrum utilization, recently the Mutli-hop Cognitive
Radio Ad-hoc Networks has gained a lot of attention and emerged as a
promising technology. In fact, there is a exponential boost in cognitive
radio network research due to the proliferation of consumer-based
applications. Nevertheless, seeing cognitive radio networks functional
on-ground yet required considerable efforts. Thus, this thesis is one
step further and we hope that it will open new horizon of research in
realizing practical cognitive radio network applications. Evidently,
dealing the cutting edge cognitive radio network research problems and
working with eminent researchers, i have gained technical experience in
terms of research. Besides this, working with SURF, i got deeper insight
into how to tackle a research problem and identify the pros and cons of
the proposed solution. For instance, (1) both PR and CR traffic should be considered because a good channel in terms of PR does not mean the channel is appropriated, and (2) for a good dissemination transmitter and receiver should be tuned to the same channels. Moreover, generally cognitive radio network
should be handled in such a way that i meet the requirements of
consumer-based applications, while considering the technological
constraints.  I hope that during the course of my PhD Thesis, i
will be able to make a remarkable contribution in the scientific community


\let\cleardoublepage\clearpage
\textheight = 650pt
\input{./chap_tex/bib_mub}


\end{document}

%% file: chap_tex/headers.tex
\if@twoside         
\def\ps@withline{\let\@mkboth\markboth
\def\@oddfoot{}\def\@evenfoot{}
\def\@evenhead{
\hbox{}\protect\rule[-1.5mm]{\textwidth}{.08mm}
\protect\hspace{-\textwidth}
\rm \thepage\hfil \large\bf \leftmark}
\def\@oddhead{
\hbox{}\protect\rule[-1.5mm]{\textwidth}{.08mm}
\protect\hspace{-\textwidth}
\hbox{}\large\bf \rightmark \hfil \rm\thepage}
\def\chaptermark##1{\markboth {
{\ifnum \c@secnumdepth >\m@ne
      \@chapapp\ \thechapter. \ \fi ##1}}{}}%
\def\sectionmark##1{\markright {
{\ifnum \c@secnumdepth >\z@
   \thesection. \ \fi ##1}}}}
\else               
\def\ps@withline{\let\@mkboth\markboth
\def\@oddfoot{}\def\@evenfoot{}
\def\@oddhead{\hbox {}\sl \rightmark \hfil \rm\thepage}
\def\chaptermark##1{\markright {\uppercase{\ifnum \c@secnumdepth >\m@ne
  \@chapapp\ \thechapter. \ \fi ##1}}}}
\fi

\def\PARstart#1#2{\begingroup\def\par{\endgraf\endgroup\lineskiplimit=0pt}
     \setbox2=\hbox{#2}\newdimen\tmpht \tmpht \ht2
     \advance\tmpht by \baselineskip\font\hhuge=cmr10 at \tmpht
     \setbox1=\hbox{{\hhuge #1}}
     \count7=\tmpht \count8=\ht1\divide\count8 by 1000 \divide\count7 by\count8
     \tmpht=.001\tmpht\multiply\tmpht by \count7\font\hhuge=cmr10 at \tmpht
     \setbox1=\hbox{{\hhuge #1}} \noindent \hangindent1.05\wd1
     \hangafter=-2 {\hskip-\hangindent \lower1\ht1\hbox{\raise1.0\ht2\copy1}%
     \kern-0\wd1}\copy2\lineskiplimit=-1000pt}

%% file: chap_tex/pagetitre-en.tex
\selectlanguage{english}
\typeout{*** English title page ***}

{
\setlength{\oddsidemargin}{0cm}
\setlength{\evensidemargin}{0cm}
\setlength{\topmargin}{0cm}
\setlength{\textwidth}{19cm}
\setlength{\textheight}{24cm}
\setlength{\headheight}{0cm}
\setlength{\headsep}{0cm}
\setlength{\parindent}{1cm}

\thispagestyle{empty}

\begin{center}
\mbox{}\\
\vspace{-2cm}
{\Large\bf Doctor of Philosophy -- Mid Thesis Activity Report \\
Pierre \& Marie Curie University \\}

\vspace{1.5cm}

{\large Specialization \\}

\vspace{0.5cm}

{\Large \bf C{\normalsize OMPUTER} S{\normalsize CIENCE} \\}

\vspace{1.5cm}

{\large presented by\\}

\vspace{0.5cm}

{\Large \bf Mr Mubashir Husain Rehmani\\}

\vspace{1.5cm} {\large Submitted as \\}

\vspace{0.5cm} {\large\bf Mid-Thesis Activity Report \\}

\vspace{1cm} {\large for the partial requirement of \\}

\vspace{0.5cm} {\large\bf Doctor of Philosophy from Pierre \& Marie Curie University \\}

\vspace{2.0cm} {\Large
  \begin{tabular}{|c|}
    \hline
    \bf Data Dissemination in Cognitive Radio Networks\\
    \hline
  \end{tabular}\\
  }

\vspace{0.5cm} {\large $24^{th}$ March 2010 \\}

\vspace{2.0cm} {\large \bf Commitee:}

 \vspace{0.5cm} {\large \bf
  \begin{tabular}{lll}
  \footnotesize{Dr. Martin May} & \footnotesize{Examiner} &  \footnotesize{Scientist and Thomson Fellow, Thomson Laboratory Paris - France}\\
  \footnotesize{Dr. Nguyen Thi Mai Trang} & \footnotesize{Examiner}  & \footnotesize{Asstt. Professor, Pierre \& Marie Curie University -- Paris - France}\\
  \footnotesize{Dr. Aline Carneiro Viana} & \footnotesize{Co-Advisor} & \footnotesize{Research Scientist, ASAP, INRIA -- Saclay - France}\\
    \footnotesize{Dr. Hicham Khalife} & \footnotesize{Collaborator} & \footnotesize{Asstt. Professor, LaBRI/ENSEIRB -- Bordeaux - France}\\
  \footnotesize{Prof. Serge Fdida} & \footnotesize{Main Advisor} &  \footnotesize{Professor, Pierre \& Marie Curie University -- Paris - France}\\

  \end{tabular}\\
  }
\end{center}
}